\documentclass{PoS}

\title{Finite temperature phase transition of two-flavor QCD with an
improved Wilson quark action}

\ShortTitle{Finite temperature phase transition of two-flavor QCD with
an improved Wilson quark action}

\author{\speaker{N.~Ukita}, S.~Ejiri, T.~Hatsuda, N.~Ishii, Y.~Maezawa\\

        Department of Physics, The University of Tokyo, \\
        Bunkyo-ku, Tokyo 113-0033, Japan

        E-mail: \email{ukita@nt.phys.s.u-tokyo.ac.jp}}

\author{S.~Aoki and K.~Kanaya \\

        Graduate School of Pure and Applied Sciences, University of Tsukuba, \\
        Tsukuba, Ibaraki 305-8571, Japan
        }

\abstract{We study the phase structure of QCD at finite
 temperatures with two flavors of dynamical quarks on a lattice
 with the size $N_s^3 \times N_t=16^3 \times 4$, using
 a renormalization group improved gauge action and a clover
 improved Wilson quark action. The simulations are made along
 the lines of constant physics determined in terms of $m_{\rm PS}/m_{\rm
 V}$ at zero-temperature. 
 We show prelimnary results for the spatial string tension in the high
temperature phase.}

\FullConference{XXIVth International Symposium on Lattice Field Theory\\
                July 23-28, 2006\\
                Tucson, Arizona, USA}

\begin{document}
\section{Introduction}
Recent relativistic heavy ion collision experiments have revealed various remarkable 
properties of QCD at finite temperatures and densities, 
suggesting the realization of the QCD phase transition from the hadronic matter to 
the quark-gluon plasma (QGP). In order to
extract an unambiguous signal for the transition from the heavy ion
collision experiments,
it is indispensable to make quantitative calculation of the 
 thermal properties of QGP  from first principles.
Currently, the lattice QCD simulation is the only systematic method to do so.
By now, most of the lattice QCD studies at finite temperature and chemical potential 
have been performed using staggered-type quarks which
 require less computational costs than others. However, 
 the lattice artifacts of the staggered-type quarks are not fully understood.
 Therefore, it is important to compare the results from 
other lattice quarks to control and estimate the lattice discretization errors.
 
We have started systematic studies of finite temperature/density QCD 
using Iwasaki's RG-improved gauge action \cite{rg} and a clover-improved
Wilson quark action \cite{cl}. 
In particular, we perform simulations along the lines of constant
physics (LCP's) to clearly extract the temperature- and density-dependences.
As the first project in this direction, we are carrying out simulations
of $N_f=2$ QCD on an  
$N_s^3 \times N_t=16^3 \times 4$ lattice 
at $m_{\rm PS}/m_{\rm V}=0.65$ and 0.80 
in the range $T/T_{pc}\sim 0.76$--3.2, where $T_{pc}$ is the
pseudocritical point along the LCP. 
Basic properties of the system at finite temperatures, such as the phase
structure and the equation of state, have been studied in
\cite{cp1,cp2}.  
In contrast to the studies with the staggered-type quarks, the expected
$O(4)$ scaling for $N_f=2$ QCD was confirmed. 
We extend the study to analyse detailed properties of QGP at finite
temperature and chemical potential.

In this paper, we report on the status of our simulations and present a
preliminary result for the spatial string tension at finite temperature.  
Results for the free energies of static quarks at zero chemical
potential are presented in \cite{maezawa}. 
Preliminary results of the equation of state and susceptibilities at
non-zero chemical potential by the Taylor expansion method are given in
\cite{ejiri}. 

\section{Lattice action}

We employ the RG-improved gauge action \cite{rg} and the
$N_f=2$ clover-improved Wilson quark action \cite{cl} defined by
\begin{eqnarray}
  &&S = S_g + S_q, \\
  &&S_g = 
  -{\beta}\sum_x\left(
   c_0\sum_{\mu<\nu;\mu,\nu=1}^{4}W_{\mu\nu}^{1\times1}(x) 
   +c_1\sum_{\mu\ne\nu;\mu,\nu=1}^{4}W_{\mu\nu}^{1\times2}(x)\right), \\
  &&  S_{q} = \sum_{f=1,2}\sum_{x,y}\bar{q}_x^f D_{x,y}q_y^f,
\end{eqnarray}
where $\beta=6/g^2$, $c_1=-0.331$, $c_0=1-8c_1$ and
\begin{eqnarray}
 D_{x,y} = \delta_{xy}
   -{K}\sum_{\mu}\{(1-\gamma_{\mu})U_{x,\mu}\delta_{x+\hat{\mu},y}
    +(1+\gamma_{\mu})U_{x,\mu}^{\dagger}\delta_{x,y+\hat{\mu}}\}
   -\delta_{xy}{c_{SW}}{K}\sum_{\mu>\nu}\sigma_{\mu\nu}F_{\mu\nu}.
\end{eqnarray}
Here $K$ is the hopping parameter, $F_{\mu\nu}$ is the lattice field
strength with $f_{\mu\nu}$ the standard clover-shaped combination of gauge links,
$F_{\mu\nu} = {1}/{8i}(f_{\mu\nu}-f^{\dagger}_{\mu\nu})$.
For the clover coefficient $c_{SW}$, we adopt a mean field value using
$W^{1\times 1}$ which was calculated in the one-loop perturbation theory
\cite{rg}, \vspace{ -5mm}
\begin{eqnarray}
 {c_{SW}}=(W^{1\times 1})^{-3/4}=(1-0.8412\beta^{-1})^{-3/4}.
\end{eqnarray}
\begin{figure}[t]
  \begin{center}
    \begin{tabular}{c}
    \includegraphics[width=70mm]{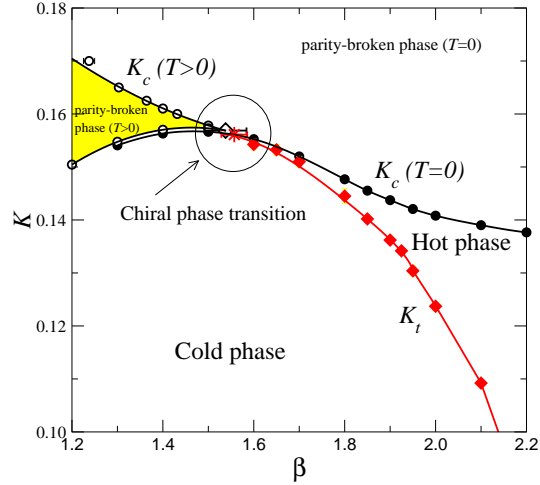}
    \end{tabular}
    \caption{Phase diagram for RG improved gauge action and clover
   improved Wilson quark action for $N_t=4$.}
    \label{fig1}
  \end{center}
\end{figure}
The phase diagram of this action in the $(\beta,K)$ plane has been
obtained by the CP-PACS 
Collaboration \cite{cp1,cp2} as shown in Fig.\ref{fig1}. The solid line $K_c(T=0)$
with filled circles is the chiral limit where pseudoscalar
mass vanishes at zero temperature. Above the $K_c(T=0)$ line, the parity-flavor
symmetry is spontaneously broken. At finite temperatures, the cusp of the
parity-broken phase retracts from the large $\beta$ limit to a finite
$\beta$. The solid line $K_c(T>0)$ connecting open symbols
represents the boundary of the parity-broken phase. The red line
$K_t$ represents the finite temperature pseudocritical line determined
from the peak of Polyakov loop susceptibility. This line separates the hot
phase (the quark-gluon plasma phase) and the cold phase (the hadron phase).
The crossing point of the $K_t$ and the $K_c(T=0)$ lines is the chiral
phase transition point.

\section{Determination of lines of constant physics and simulation parameters}

\begin{figure}[t]
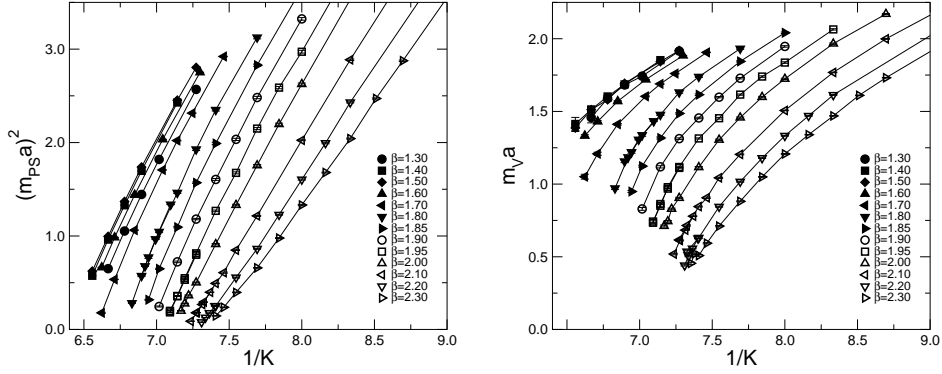

  \begin{center}
    \begin{tabular}{cc}
    \includegraphics[width=60mm]{Fig/pi2-k_0607012.eps} &
    \includegraphics[width=60mm]{Fig/rho-k_060712p.eps}
    \end{tabular}
    \caption{Pseudoscalar meson mass squared (left) and vector meson
   mass (right) as a function of $1/K$ for several values of $\beta$ at $T=0$.}
    \label{fig2}
  \end{center}
\end{figure}

\begin{figure}[t]
  \begin{center}
    \begin{tabular}{c}
    \includegraphics[width=80mm]{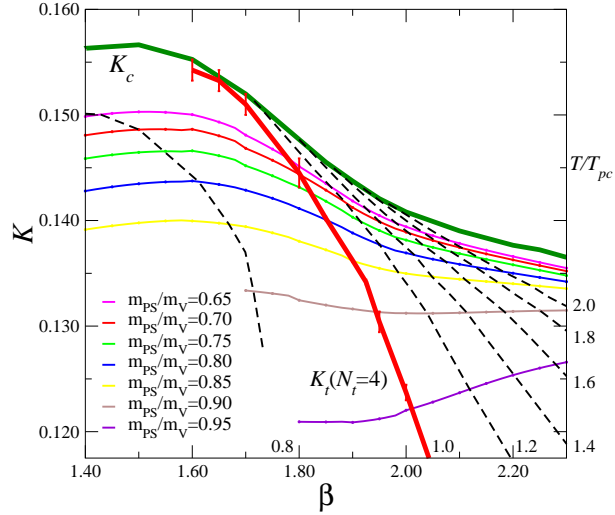}
    \end{tabular}
    \caption{Lines of constant $m_{\rm PS}/m_{\rm V}$ and constant
   $T/T_{ps}$ in the $(\beta,K)$ plane.}
    \label{fig3}
  \end{center}
\end{figure}

For phenomenological applications, we need to investigate the
temperature dependence of thermodynamic observables on a line of
constant physics (LCP), which we determine by the ratio $m_{\rm PS}/m_{\rm V}$ 
of pseudoscalar and vector meson masses. 
For our purpose, we need LCP in a wider range of parameters than \cite{cp2}. 
Therefore, we re-analyze the data for $m_{\rm PS}a$ and $m_{\rm V}a$ at
zero temperature shown in Fig.\ref{fig2} determined
by the CP-PACS Collaboration \cite{cp1,cp2,cp3,cp4}. 
The colored solid lines in Fig.\ref{fig3} shows our results for LCP
corresponding to $m_{\rm PS}/m_{\rm V}=0.65, 0.70, 0.75, 0.80, 0.85,
0.90$ and $0.95$. The green line denoted as $K_c$ represents the
critical line, i.e. $m_{\rm PS}/m_{\rm V}=0$.
Our LCP's are consistent with those of \cite{cp2} in the range of
the previous study.


We also reanalyze the lines of constant temperature $T/T_{pc}$.
The temperature $T$ is estimated by the zero-temperature vector meson
mass $m_{\rm V}a(\beta,K)$ using
\begin{eqnarray}
 \frac{T}{m_{\rm V}}(\beta,K)=\frac{1}{N_t \times m_{\rm V}a(\beta,K)}.
\end{eqnarray}
The lines of constant $T/T_{pc}$ is determined as the ratio of
${T}/{m_{\rm V}}$ to ${T_{pc}}/{m_{\rm V}}$ where
$T_{pc}/m_{\rm V}$ is obtained by $T/m_{\rm V}$ at $K_t$ on the same LCP.
The bold red line denoted as $K_t(N_t=4)$ in Fig.\ref{fig3} represents the
pseudocritical line $T/T_{pc}=1$.
The dashed lines represent the results for 
$T/T_{pc}=0.8, 1.2, 1.4, 1.6, 1.8, 2.0$ at $N_t=4$.

We perform finite temperature simulations on a lattice with a
temporal extent $N_t=4$ and a spatial extent $N_s=16$ along the LCP's for 
$m_{\rm PS}/m_{\rm V}=0.65$ and $0.80$. 
The standard hybrid Monte Carlo algorithm is employed to generate full QCD
configurations with two flavors of dynamical quarks. The length of one
trajectory is unity and the molecular dynamics step size is tuned to
achieve an acceptance rate greater than about 70\%.
Runs are carried out in the range $\beta=1.50$--2.30 at twelve values of
$T/T_{pc}\sim 0.82-3.2$ for $m_{\rm PS}/m_{\rm V}=0.65$ and eleven values of
$T/T_{pc}\sim 0.76-2.5$ for $m_{\rm PS}/m_{\rm V}=0.80$.
Our simulation
parameters and the corresponding temperatures are summarized in Table 1.
The number of trajectories for each run after thermalization is
$5000-6000$. We measure physical quantities at every 10 trajectories.

\begin{table}
\caption{Simulation parameters for $m_{\rm PS}/m_{\rm V}=0.80$ (left) 
and $m_{\rm PS}/m_{\rm V}=0.65$ (right).}
\begin{center}
\begin{tabular}[t]{cclcc}\hline
    $\beta$  & $K$ & $T/T_{pc}$   & Traj. \\ \hline
       1.50  & 0.143480 & 0.76446 & 5500\\ \hline
       1.60  & 0.143749 & 0.79544 & 6000\\ \hline
       1.70  & 0.142871 & 0.84346 & 6000\\ \hline
       1.80  & 0.141139 & 0.92507 & 6000\\ \hline
       1.85  & 0.140070 & 0.98642 & 6000\\ \hline
       1.90  & 0.138817 & 1.07619 & 6000\\ \hline
       1.95  & 0.137716 & 1.19836 & 6000\\ \hline
       2.00  & 0.136931 & 1.34778 & 5000\\ \hline
       2.10  & 0.135860 & 1.69025 & 5000\\ \hline
       2.20  & 0.135010 & 2.07325 & 5000\\ \hline
       2.30  & 0.134194 & 2.51093 & 5000\\ \hline
\end{tabular}
\hspace{1cm}
\begin{tabular}[t]{cclcc}\hline
  $\beta$& $K$ & $T/T_{pc}$   &  Traj.\\ \hline
   1.50  & 0.150290 & 0.82434 & 5000\\ \hline
   1.60  & 0.150030 & 0.86471 & 5000\\ \hline
   1.70  & 0.148086 & 0.94442 & 5000\\ \hline
   1.75  & 0.146763 & 1.00024 & 5000\\ \hline
   1.80  & 0.145127 & 1.07466 & 5000\\ \hline
   1.85  & 0.143502 & 1.17857 & 5000\\ \hline
   1.90  & 0.141849 & 1.31675 & 5000\\ \hline
   1.95  & 0.140472 & 1.48262 & 5000\\ \hline
   2.00  & 0.139411 & 1.66828 & 5000\\ \hline
   2.10  & 0.137833 & 2.09054 & 5000\\ \hline
   2.20  & 0.136596 & 2.59279 & 5000\\ \hline
   2.30  & 0.135492 & 3.21536 & 5000\\ \hline 
\end{tabular}
\end{center}
\end{table}
\section{Spatial Wilson Loop}

Using the stored configurations, we are carrying out a series of studies
on the nature of the quark-gluon plasma.
In this report, we present our preliminary results on the confinement in
the spatial directions at high temperature.

In previous quenched studies\cite{bali,karr,karsch,boyd},
Wilson loops in spatial directions are
found to show non-vanishing spatial string tension $\sigma_s$
even at $T>T_{pc}$, which is called the spatial confinement.
We study this phenomenon when there exist dynamical quarks in the system.
Altough quarks are expected to decouple from the
spatial observables for $T \gg T_{pc}$ due to dimensional reduction
and thus do not affect $\sigma_s$ in the high temperature limit,
it is not obvious whether the same is true near $T_{pc}$.

\begin{figure}[t]
  \begin{center}
    \begin{tabular}{cc}\hspace{-10mm}
    \includegraphics[width=65mm]{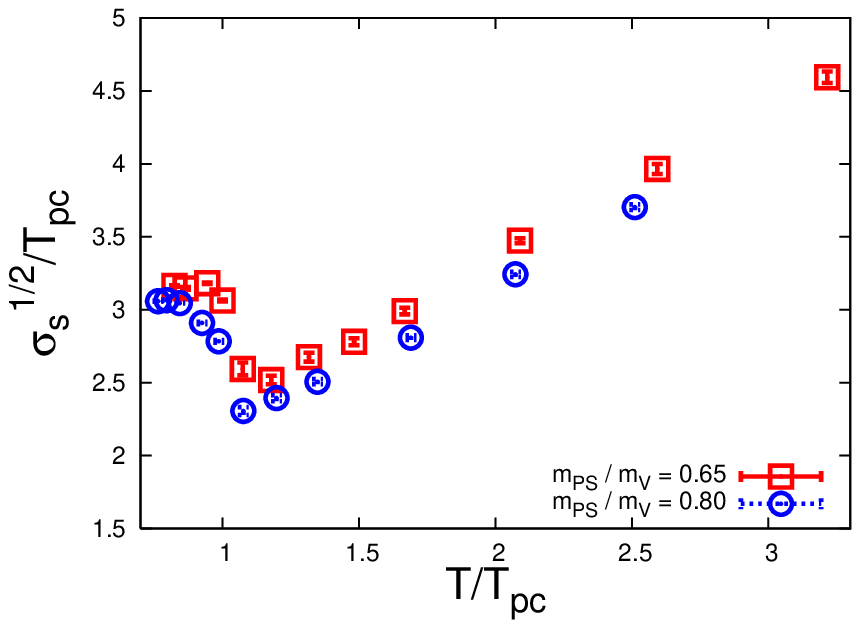} &
    \includegraphics[width=65mm]{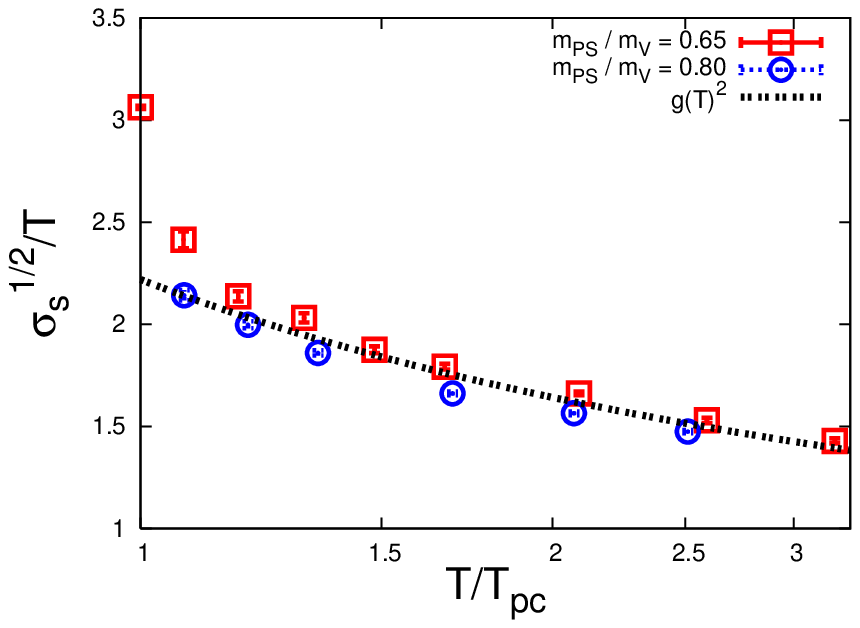}
    \end{tabular}
    \caption{The square root of the spatial string tension over $T_{pc}$ (left) and $T$
   (right) as a function of $T/T_{pc}$ for $m_{\rm PS}/m_{rm V}=0.65,
   0.80$. The dashed line (right) shows the two-loop fit.}
    \label{fig4}
  \end{center}
\end{figure}

As a first trial, we evaluate $\sigma_s$ assuming the simplest ansatz
for the spatial Wilson loops
$W(I,J)$ with the size $I\times J $:
\begin{eqnarray}
 -\ln W(I,J) = \sigma_s IJ + \sigma_{peri}(2I+2J) + C_w,
\end{eqnarray}
where $\sigma_s$, $\sigma_{peri}$ and $C_w$ are fit parameters.
The results for $\sqrt{\sigma_s(T)}/T_{pc}$ are shown in Fig.\ref{fig4}
as a function of $T/T_{pc}$.
We find that $\sqrt{\sigma_s(T)}/T_{pc}$ tends to a non-vanishing value below
$T_{pc}$ while it increases linearly in $T/T_{pc}$ above $T_{pc}$.
Similar behavior was observed in the quenched case too \cite{bali}.

Let us now compare the results with a calculation of  $\sqrt{\sigma_s}$
by the three-dimensional effective theory assuming the dominance of the gauge part.
We would expect the following behaviour for the spatial string tension,
\begin{eqnarray}
 {\sqrt{\sigma_s(T)}} &=& c\ g^2(T)\ T,
\end{eqnarray}
where  $g^2(T)$ is the two-loop temperature
dependent running coupling constant in four dimensions,
\begin{eqnarray}
 g^{-2}(T) = \frac{29}{48\pi^2}\ln\frac{T}{\Lambda}
         + \frac{226}{768\pi^4}\ln\left(2\ln\frac{T}{\Lambda}\right).
\end{eqnarray}
We carry out a fit to our data shown in Fig.\ref{fig4},  regarding $c$ and $\Lambda$ free
parameters.
From the two parameter fit, we find \hspace{-10mm}
\begin{eqnarray}
 c=0.54(6), \ \ \Lambda/T_{pc} = 0.14(4).
\end{eqnarray}
We note that these values are similar to those obtained in a quenched
study \cite{karsch,boyd}:
$c=0.566(13)$ and $\Lambda/T_{pc} = 0.104(9)$.

\begin{figure}[t]
  \begin{center}
    \begin{tabular}{cc}
    \includegraphics[width=65mm]{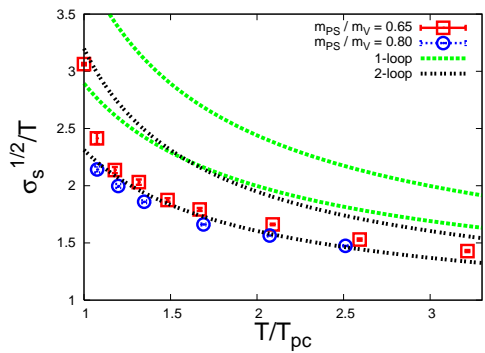}
    \end{tabular}
    \caption{Comparison of our data and 1-loop and 2-loop predictions
   from three-dimensional effective theory. $\overline{\mu}/T$ is fixed
   as in \cite{ls}. $T_{pc}/\Lambda_{\overline{\rm MS}}$ is varied in
   the range $0.53-0.77$.} 
    \label{fig5}
  \end{center}
\end{figure}

We also note that there is a parameter-free three-dimensional effective
theory prediction for 
$\sqrt{\sigma_s(T/T_{pc},T_{pc}/\Lambda_{\overline{\rm MS}},\overline{\mu}/T)}$
\cite{ls}, where  $\overline{\mu}$ is the $\overline{{\rm MS}}$ scheme
scale parameter. The $\overline{\mu}/T$ is fixed as in \cite{ls}. 
We vary the value of $T_{pc}/\Lambda_{\overline{\rm MS}}$
in the range $0.53-0.77$ \cite{cp1,g}.
The results for $\sqrt{\sigma_s} / T$ are compared with our data in
Fig.\ref{fig5}. While the 1-loop prediction of the effective
theory (shown by the region bounded by two green lines in Fig.\ref{fig5})
deviates from the lattice data, the 2-loop result (shown by the region
bounded by two black lines) is roughly consistent with our data. A more
detailed test is left for a future work.

\section{Conclusions}

Most of the lattice QCD studies at finite temperatures and densities have been  
done using staggered-type quarks.
To control the lattice artifacts, comparisons with other lattice quarks
are indispensable. 
Therefore, we have started a systematic study of QCD at finite temperature and
density 
with an improved Wilson quark action.

As a first step, we performed simulations of $N_f=2$ QCD on an 
$N_s^3 \times N_t=16^3 \times 4$ lattice.
We have identified the lines of constant physics and studied the
temperature-dependence of
various quantities at $m_{\rm PS}/m_{\rm V}=0.65$ and 0.80
in the range $T/T_{pc}\sim 0.76$--3.2.
 A preliminary result of the spatial string tension $\sigma_s(T)$ 
 in the quark-gluon plasma
 shows a behavior consistent with  ${\sqrt{\sigma_s(T)}} = c\ g^2(T)\ T$
 where $c$ takes a value close to that in the quenched case (gluon plasma).
  $\sigma_s(T)$  is also consistent with the  
 parameter-free prediction of the three-dimensional effective theory
 in the 2-loop order. 
 Further results on the static quark free energies at finite temperatures
for different color channels are given in \cite{maezawa}.
Results on the equation of state and susceptibilities at non-zero chemical potential with
Wilson-type quarks are shown in \cite{ejiri}.

\paragraph{Acknowledgements:}
This work is supported by Grants-in-Aid of the Japanese
 Ministry of Education, Culture, Sports, Science and Technology,
(Nos. 13135204, 15540251, 17340066, 18540253, 18740134).
SE is supported by Sumitomo Foundation (No. 050408), and
YM is supported by JSPS.
This work is in part supported also by the Large-Scale Numerical
Simulation Projects of ACCC, Univ. of Tsukuba, and by the Large Scal
Simulation Program of High Energy Accelerator Research Organization
(KEK).


\begin{thebibliography}{99}
  \bibitem{rg} Y. Iwasaki, Nucl. Phys. B {\bf 258}, 141 (1985);
          University of Tsukuba Report No. UTHEP-118 (1983).

  \bibitem{cl} B. Sheikholeslami and R. Wohlert, Nucl. Phys. B {\bf 259}, 572 (1985).

   \bibitem{cp1} CP-PACS Collaboration, A. Ali Khan {\it et al.},
          Phys. Rev. D {\bf 63}, 034502 (2000).

  \bibitem{cp2} CP-PACS Collaboration, A. Ali Khan {\it et al.},
          Phys. Rev. D {\bf 64}, 074510 (2001).

  \bibitem{maezawa} Y. Maezawa {\it et al.}, PoS {\bf LAT2006}, 141.

  \bibitem{ejiri} S. Ejiri {\it et al.}, PoS {\bf LAT2006}, 132.

  \bibitem{cp3} CP-PACS Collaboration, A. Ali Khan {\it et al.},
          Phys. Rev. Lett. {\bf 85}, 4674 (2000).

  \bibitem{cp4} CP-PACS Collaboration, A. Ali Khan {\it et al.},
          Phys. Rev. D {\bf 65}, 054505 (2001).

  \bibitem{bali} G. S. Bali {\it et al.}, Phys. Rev. Lett. {\bf 71}, 3059
          (1993).

  \bibitem{karr} L. K\"{a}rkk\"{a}inen {\it et al.}, Phys. Lett. B {\bf 312}, 173 (1993).

  \bibitem{karsch} F. Karsch {\it et al.}, Phys. Lett. B {\bf 346}, 94 (1995).

  \bibitem{boyd} G. Boyd {\it et al.}, Nucl. Phys. B {\bf 469}, 419 (1996).

  \bibitem{ls} Y. Schr\"{o}der and M. Laine, PoS {\bf LAT2005}, 180 (2006).

  \bibitem{g} M. Gockeler {\it et al.},  Phys. Rev. D {\bf 73}, 04506
	  (2004).

\end{thebibliography}
\end{document}